\documentclass[%
 reprint,
superscriptaddress,
nofootinbib,
 amsmath,amssymb,
 aps,
]{revtex4-2}
\usepackage[T1]{fontenc}
\usepackage{color}
\usepackage{graphicx}
\usepackage{subfig}
\usepackage{xcolor}
\usepackage{upgreek}
\usepackage{tabularx}
\usepackage{ulem}
\usepackage{comment}
\usepackage{chemist}
\usepackage{chemmacros}

\begin{document}
\title{Sol-gel transition by evaporation in porous media}

\author{Romane Le Dizès Castell}
    \email{r.ledizes2@uva.nl}
\affiliation{Institute of Physics, University of Amsterdam, The Netherlands}
\author{Leo Pel}
\affiliation{Department of Applied Physics, Eindhoven University of Technology, The Netherlands}
\author{Tinhinane Chekai}
\affiliation{Universite de Pau et des Pays de l’Adour, E2S UPPA, CNRS, TotalEnergies, LFCR, Pau, France}
\affiliation{Universite de Pau et des Pays de l’Adour, E2S UPPA, CNRS, DMEX, Pau, France}
\author{Hannelore Derluyn}
\affiliation{Universite de Pau et des Pays de l’Adour, E2S UPPA, CNRS, TotalEnergies, LFCR, Pau, France}
\affiliation{Universite de Pau et des Pays de l’Adour, E2S UPPA, CNRS, DMEX, Pau, France}
\author{Mario Scheel}
\affiliation{Synchrotron Soleil, Saint-Aubin, France}  
\author{Sara Jabbari-Farouji}
\affiliation{Institute of Physics, University of Amsterdam, The Netherlands}
\author{Noushine Shahidzadeh}
\affiliation{Institute of Physics, University of Amsterdam, The Netherlands}

\begin{abstract}
Historical monuments, outdoor stone sculptures and artworks made of porous materials are exposed to chemical and physical degradations over time. Presently, the most promising route for consolidation of weakened porous materials is the injection of viscoelastic solutions of polymerizing compounds. Those compounds, after injection, undergo a sol-gel transition inside the porous media through evaporation of the solvent. Finding a suitable gelifying solution as a consolidant calls for understanding the drying kinetics of viscoelastic fluids in porous media. Here, we present a multiscale study of the drying kinetics of fluids during the sol-gel transition in porous materials using NMR and X-ray microtomography techniques. We find that from the early stage of the drying, a heterogeneous desaturation develops and advances from the free surface of evaporation towards the inner parts of the stone. We identify different drying periods which appear to be dependent on the intrinsic properties of the porous medium influencing strongly the homogeneity of the final gel distribution within a treated stone. Our findings not only are relevant for the consolidation of porous artworks but also for civil and soil engineering processes where the fluids considered are generally more complex than water.
\end{abstract}

\maketitle

\section{Introduction}
Artworks (such as sculptures, statues, frescos or tiles) made of porous materials (sandstone, carbonate stone, terracotta and clays) are frequently exposed to environmental fluctuations, \textit{i.e.} changes in relative humidity and temperature leading to freeze-thaw \cite{al-omari_effect_2014, bayram_predicting_2012}, salt crystallization and swelling/deswelling \cite{shahidzadeh-bonn_damage_2010, gibson_characterisation_1997, rijniers_salt_2005, siegesmund_natural_2002, espinosa-marzal_impact_2013}. These effects are widely accepted to be the main causes of the long-term material degradation of stone artworks located either indoors or outdoors \cite{shahidzadeh-bonn_damage_2010, gibson_characterisation_1997, hu_atmospheric_2009, linnow_analysis_2007}. Indeed, they reduce the cohesion of the granular matrix and consequently the mechanical integrity of the object which manifest itself in the form of flaking of the surface layers, powdering and granular disintegration \cite{desarnaud_pressure_2016, flatt_chemo-mechanics_2014}. To preserve this weakened materials, consolidation treatments are therefore necessary. In most of the cases, decisions to intervene are made based on the experience of the conservation scientists by employing a limited array of consolidating and protective treatments. \\

Fluids undergoing a sol-gel transition are often used to consolidate mechanically weakened materials \cite{wheeler_alkoxysilanes_2005, artesani_recent_2020}. They are injected in the porous network of the materials and after the capillary impregnation, consolidation occurs with the solvent evaporation followed by chemical reactions leading to the formation of films of a new material adhering to the grains of the porous network and reinforcing their cohesion. The starting sol is a Newtonian liquid which with solvent evaporation and condensation reactions is converted into a viscoplastic material. This is characterized by an abrupt increase of the viscosity and the formation of a gel network. Gels obtained from tetraethoxysilane (TEOS) as precursor are commonly to reinforce sandstones artworks as they can effectively bond to the silicate group of the stone (quartzic) due to their chemical affinities \cite{potzl_consolidation_2021, wheeler_alkoxysilanes_2005, tsakalof_assessment_2007, artesani_recent_2020}. Nevertheless, TEOS gels have also been reported to crack due to the strong capillary stresses which can develops during the evaporation of the solvent making the final treatment inefficient \cite{mosquera_stress_2003, mosquera_producing_2003}. To get around this problem and reduce the capillary stresses, surfactants \cite{mosquera_surfactant-synthesized_2010} or silica nanoparticles have been proposed as additives to the TEOS suspension \cite{kim_effects_2009,liu_preparation_2013}.

Generally, an “ideal” consolidation treatment should improve the mechanical resistance of the porous objects with minimal changes of their other properties such as appearance, porosity or
permeability. A homogeneous distribution of the treatment is also important as fragile materials usually break at their weakest points. In order to ensure homogeneity, the dynamics of evaporation and gelation in the confined porous network should be studied. Although the physics of the evaporation of various Newtonian fluids in porous media (suspended particles \cite{keita_mri_2013, keita_particle_2021, qin_lbm_2019, qin_lattice_2023}, salt solutions \cite{desarnaud_drying_2015, shahidzadeh-bonn_salt_2008, veran-tissoires_salt_2012}, etc) are quite well reported in the literature, the case where a suspension converts to a non-Newtonian fluids during evaporation is still poorly understood. There is still a lack of in-depth knowledge about the drying kinetics leading to the sol- gel transition and the relation to the effectiveness of such treatments during their application. \\

Here, we report a multiscale study on the evaporation followed by sol-gel transition of methytriethoxysilane (MTEOS) suspensions in sandstone samples and its consequences on the drying kinetics. Our aim is to relate the drying kinetics and the final distribution of gel in the porous medium to the gelation kinetics. To this end, the remainder of this article is organised as follows.  The drying of methyltriethoxysilane in sandstone is first studied with NMR technique during the gelation process. The distribution of the sol and gel during and at the end of the drying is then investigated using computed X-ray microtomography. Our results show that using methytriethoxysilane (MTEOS) as precursor can lead the formation of a crack-free gel without any additives as the free methyl group reduces the capillary pressure during condensation and polymerization. We also show that the gel mainly forms in small pores near the evaporative surface. Finally, we propose an extensive description of the mechanism of gel formation during evaporation in the porous medium based on our measurements. Our findings provide new insights into the gelation dynamics with evaporation in porous media and help for taking a better conservation measures depending on the materials and environmental factors.  More generally, the present article contributes to a better understanding of drying with complex fluids in porous media, which is relevant for numerous physical and industrial situations.
\\


\section{Methods}

\subsection{Preparation of the sol and description of the sol-gel process}
\begin{figure*}[http]
    \centering
    \includegraphics[width=0.99\textwidth]{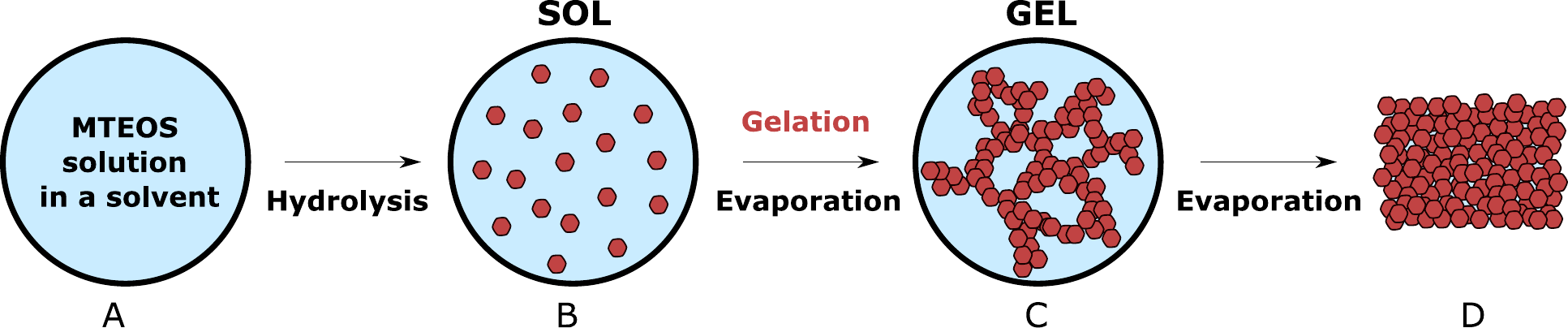}
    \caption{Schematic of different stages of the sol-gel transition by evaporation. (A) MTEOS is in solution in acidic water. (B) The hydrolysed MTS polymerise to form oligomers, that grow into macromolecules (MTS) of radius 2-3nm. Those macromolecules are thus in suspension in the solvent and form the sol. (C)  With condensation reaction and evaporation of the solvent, a gel network is formed, spanning throughout the whole sample. (D) Evaporation of most of the solvent and densification of the gel. Some liquid remains stuck in the nanopores of the gel and does not evaporate.}
    \label{solgel}
\end{figure*}

Methyltriethoxysilane \chemform{(C_2H_5O)_3SiCH_3} (Sigma-Aldrich) is dissolved in 0.1 mol/L solution of acetic acid (Sigma-Aldrich). The concentration of MTEOS is 2.5 mol/L in the acidic solution. Upon hydrolysis, silanol groups are formed and ethanol is released in the solution based on the following reaction : 
\begin{reactions}
 (CH3CH2O)3SiCH3 + $3$ H2O -> \\ 
CH3Si(OH)3 + $3$ CH_3CH_2OH, \notag 
\end{reactions}
The resulting solution when completely hydrolyzed contains \chemform{CH_3Si(OH)_3} methylsilanetriol (MTS) in a solvent composed of water and ethanol. Generally, condensation reactions between the formed MTS monomers start before the hydrolysis is complete. Through condensation, MTS monomers react with each other to create siloxane bridges and form macromolecules. The resulting suspension of macromolecules forms the \textbf{sol} (also called MTS sol in the remainder of the article), which corresponds to stage B in Fig \ref{solgel}. In acidic conditions, the macromolecules remain small in size and their growth stops when they reach a radius of 2-4 nm \cite{Iler1571135649987057536}. 
Two condensation reactions can describe the gelation which lead to the creation of (3D)-network structure (stage C in Fig \ref{solgel}) \cite{Iler1571135649987057536, brinker_sol-gel_2013} : 

\begin{reactions}
 CH_3R'_2Si - OCH_2CH_3 + HO-SiCH_3R'_2 -> \\
  CH_3R'_2Si - O - SiCH_3R'_2 + CH_3CH_2OH     \notag
\end{reactions}

\begin{reactions}
CH_3R'_2Si - OH + HO-SiCH_3R'_2 -> \\
  CH_3R'_2Si - O - SiCH_3R'_2 + H_2O  \notag
\end{reactions}
With MTS, one of the bonds of the silanol is the methyl group which does not contribute to the formation of siloxane bridges. The latter gives a more pronounced hydrophobic property to the gel and we expect that the capillary pressure which develops will be smaller than for gels obtained with TEOS as precursor. 
The condensation reactions are concurrent to the evaporation of the solvent and lead finally to the formation of a highly branched gel network spanning through the whole sample (stage D in Fig \ref{solgel}). Bond formation nonetheless does not stop at the gel point and condensation reactions and evaporation continue. The transition to the gel state is characterized by an abrupt rise of the viscosity (which reaches $10^5$ Pa.s when the sol is fully gelated) and the emergence of an elastic response to stress \cite{brinker_sol-gel_2013} (see the Appendix A for rheological measurements).

\begin{figure*}[http]
    \centering
    \includegraphics[width=0.99\textwidth]{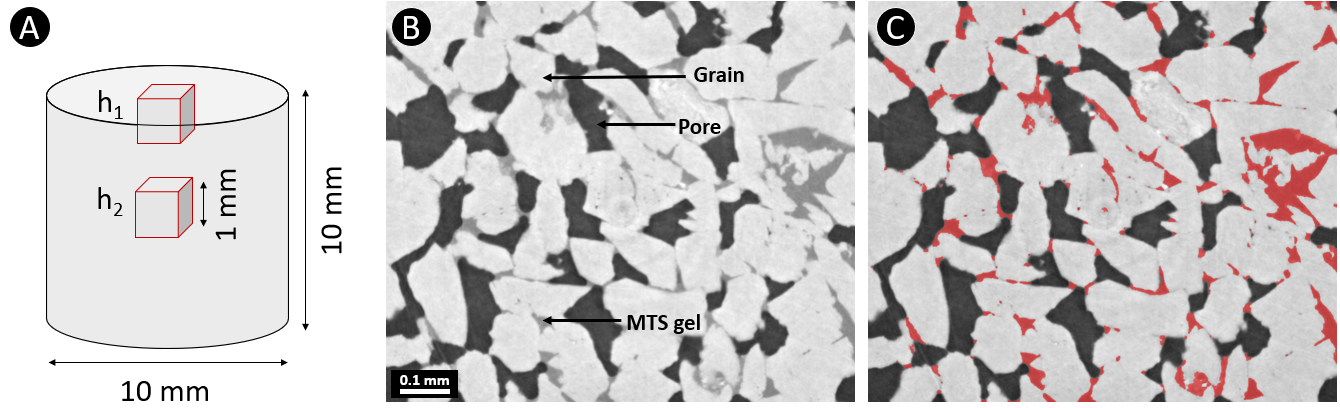}
    \caption{(A) Schematics of the stone and the positions of the scans: $h_1$ at the evaporative surface and $h_2$ at a depth of roughly 4 mm inside the stone. The reconstructed scans are cubes of volume $0.92$ mm$^3$ (B) Vertical cross section of a computed microtomography image where the fluid, the pores and the grains can be distinguished, the detection of the fluid (red-shaded areas) using segmentation on gray levels is visible in (C).}
    \label{PSStomodetect}
\end{figure*}

\subsection{Unidirectional drying experiments} Mšené sandstone (Prague, Czech Republic) composed of quartz grains of an average size around $100 \mu$m \cite{desarnaud_laboratory_2019} with a porosity of $30\%$ and an average pore diameter of $30$ $\mu$m is used in this study \cite{desarnaud_impact_2013,pavlik_water_2008}. Cylindrical samples of dimensions (height $h \times $ diameter $d$) $10$ x $10$ mm and $30$ x $20$ mm are saturated with the \textbf{sol} (\textit{i.e.} the colloidal suspension of MTS macromolecules of size 2-4nm, stage B in Fig \ref{solgel}) by spontaneous capillary rise. We study the transition from the sol to the gel state in the samples upon evaporation. Investing the process in different sample sizes gives information on the impact of the volume to surface ratio of the sample on the gelation kinetics during evaporation. 
The saturated stone samples are placed in a homemade controlled climatic chamber at $T=21^{\circ}$C and $RH=50\%$ (using the humidity regulator developped by \cite{boulogne_cheap_2019}) and dried from top surface by covering all other sides with aluminium tape. The drying kinetics are investigated using an automated balance (precision of $0.001$ g).

\subsection{NMR experiments}
The \textbf{sol} distribution within the stone sample while drying is quantified by measuring the saturation profiles using a Nuclear Magnetic Resonance set-up, the basic principle of which is briefly explained hereafter. It is important to note that due to the short relaxation time of the hydrogens in the gel, the NMR set-up measures only the hydrogens in the solvent (liquid state). The hydrogens present in the MTS macromolecules or in the gel do not intervene in the quantification with this technique.

When nuclei (such as H) possessing a magnetic moment are placed in a constant magnetic field $B_{main}$, their magnetic moment precess at a certain resonance frequency $f$ called the Larmor frequency. The resonance condition allows the NMR method to be used to recognize certain nuclei. An oscillating magnetic field introduced by a Radio Frequency (RF) pulse at the Larmor frequency of hydrogen nuclei is applied. By using a Hahn RF pulse sequence, it is possible to generate spin-echo signals \cite{hahn_spin_1950}, where the intensity of the signal $S$ is proportional to the density of hydrogen nuclei excited $\rho$ and therefore to the amount of for example moisture present: 
\begin{equation}
    S=\rho \exp (-\frac{\text{TE}}{T_2})(1-\exp (-\frac{\text{TR}}{T_1})).
\end{equation}
Here, TE is the spin echo relaxation time, $T_2$ is the spin-spin relaxation time, TR the repetition time and $T_1$ is the spin-lattice relaxation time. The NMR set-up used in our experiment is composed of an iron-cored electromagnet providing a static magnetic field $B_{main}$ of 0.8 T. A constant magnetic field gradient $B_{gradient}$ of 0.3 T/m, resulting in a longitudinal resolution in the order of 1.0 mm is used. In all the measurements, the spin-echo time TE of 250 $\mu$s and repetition time TR of 5 s are used. 

Similarly to gravimetric experiments, measurements are done with the same cylindrical samples ( $h=10$ mm $\times $ $d=10$ mm) and ($h = 30$ mm $\times $ $d=20$ mm). The saturated samples with the \textbf{sol} are placed in a teflon holder adapted to the size of the stones and suspended in the NMR set-up. Teflon is used instead of aluminium tape to confine the samples as it gives no background signal during the measurements. 
A dry airflow at the rate of 1 L/min with a relative humidity of $50\%$ is blown over the sample to create drying conditions. With a stepper motor, the sample holder can be moved along the NMR insert, allowing the moisture content to be measured over time as a function of the sample height. 

\subsection{X-ray microtomography} 
X-ray microtomography experiments are performed on the ANATOMIX beamline of Synchrotron SOLEIL in Saint-Aubin \cite{weitkamp_microtomography_2022}. The filtered white beam mean X-ray photons energy is 36 keV. The X-ray photones are collected from a detector (LuAG:Ce scintillator of 20 µm thickness, 10x Mitutoyo objective (Na 0.28), camera Hamamatsu Orca Flash 4.0 V2) resulting in an effective pixel size of 0.65 $\mu$m. The sample detector distance is 6 cm, and 2000 projections are collected over an angle of 180°. The exposure time is 80 ms resulting in a scan duration of about 3 min. Volumes are reconstructed with PyHST2 \cite{mirone_pyhst2_2014} (ESRF, Grenoble, France) and a paganin filter is applied during the data reconstruction \cite{paganin_simultaneous_2002}. 

Cylindrical sandstone samples ($h=10$ mm $\times$ $d=10$ mm) were saturated with the \textbf{sol} and covered with aluminium tape and dried in one direction in the same conditions $RH=50\%$ , $T=21^{\circ}C$. Scans are taken, every hour during drying at two different heights of the sample: $h_1$ at just beneath the evaporative surface of the stone and $h_2$ at a depth of 4 mm in the porous material (see Fig \ref{PSStomodetect}-A). 
The reconstructed scans are analysed using the software Dragonfly ORS \cite{dragonfly}. During that stage, the scans are cropped to cubes of total volume $0.92$ mm$^3$. The gray values of the computed tomography images reflect the density and atomic number of the components present in the specimen. During the drying, the image quality allows to distinguish the MTS fluid from the pore space and the grains. By a simple segmentation through manual thresholding on gray levels we are able to efficiently differentiate between fluid (sol and/or gel), pores and stone grains (see Fig \ref{PSStomodetect}-B and C). The fluid, in darker grey than the grains of the stone is clearly visible on those cross sections. 
To assess the accuracy of the segmentation, the porosity of the stone is determined from the scans by measuring the volume of the fluid and the empty pores leading to $\phi_{tomo} = 28\%$ which is in very good agreement with the porosity of the stones obtained by weight measurement and Mercury intrusion porosimetry. Nonetheless, even though the density of the fluid increases with condensation, it is not possible with this technique to distinguish the sol from the gel state of the fluid with evaporation.

The detection of the MTS fluid over the whole volume of scanned cubes (0.92 mm$^3$) allows to determine the saturation in MTS defined as $s = V_{fluid}/V_{pores}$ at different positions inside the stone. Note that the $s$ obtained by this method is different from the relative moisture content determined by NMR technique. In the latter, only the evolution of the liquid state is quantified in the pores. In the first two hours of the drying, due to the fast evaporation of the solvent, the liquid moves inside the pores, which results in 'blurry' artefact in the tomography scans (see Fig \ref{Tomokinetics}-A $t=1.5$ h and $h_2=4$ mm). This artefact could affect the accuracy of the detection, hence the estimation of sol content at early stages of drying.

\subsection{SEM} 
The dried samples containing the gel have been analysed by a Hitachi TM-3000 tabletop Scanning Electron Microscope, and compared with the results obtained by X-ray microtomography at the late stage of evaporation. This SEM operates under a low vacuum, where a gas under a variable pressure is let into the sample chamber allowing imaging of the samples without any preparation or coating on the samples. 


\section{Results and Discussion} 
 
\subsection{Drying kinetic during gel formation in porous media by NMR}
\begin{figure*}
    \centering
    \includegraphics[width=0.99\textwidth]{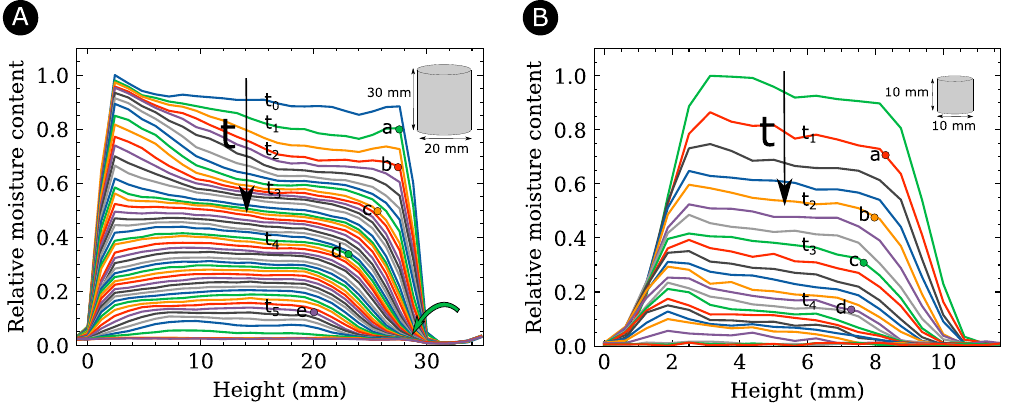}
    
    \caption{NMR profiles during unidirectional drying of the sol in a (A) big cylinder (obtained every 3 hours) and (B) small cylinder (obtained every hour). In the profiles at $t_0$ to $t_5$ some interesting points are highlighted (a to e).}
    \label{NMRSmallBig}
\end{figure*}

\begin{figure*}
    \centering
    \includegraphics[width=0.99\textwidth]{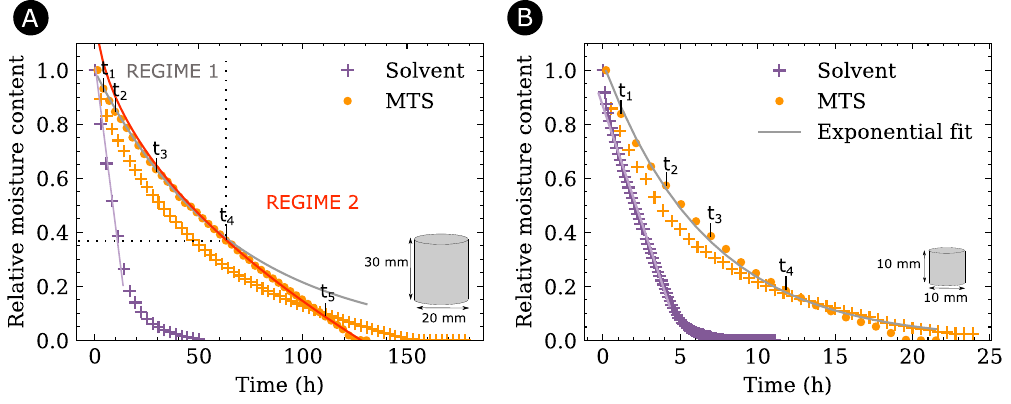}
    \caption{The mean relative liquid content as a function of time for big (A) and small (B) stones obtained by integration of the saturation profiles from NMR (dots) and weight (+) measurements. The curves are fitted by an exponential (grey line) for small stones ($y = \text{exp}(-0.15 x)$) and for big stones at short times (regime 1: $y = \text{exp}(-0.016 x)$). This is followed by a square root of time decay (red line) for big samples ($y=-0.11 \sqrt{x} + 1.2$). For the pure solvent in both size samples, the drying occurs at a constant rate ($-0.05$ h$^{-1}$ for big stones and $-0.2$ h$^{-1}$ for small stones) until very low saturation.}
    \label{NMRcurve}
\end{figure*}

The temporal evolution of the relative liquid saturation obtained by gravimetric experiments and NMR are presented in Fig 2 and Fig 3. The NMR saturation profiles, allow us to investigate the distribution of the hydrogens in the solvent over time and with the height of the sample. We recall here that the hydrogens present in the MTS macromolecules or in the gel are not detected with the NMR set-up. 

For both sample sizes, in Fig \ref{NMRSmallBig}-A and B a heterogeneous liquid content as a function of the height of the sample can be observed since the initial stage of evaporation. The moisture content is larger at the bottom of the sample (height=0mm) than at the top where evaporation takes place (height =30mm in Fig \ref{NMRSmallBig}-A and height=10 mm in Fig \ref{NMRSmallBig}-B). At the early stage of drying, this saturation gradient is due to the presence of the airflow in the NMR setup. From $t_0$ to $t_2$, in Fig \ref{NMRSmallBig}-A, although the overall saturation profiles decrease gradually with drying, the fluid remains in contact with the free surface (the height corresponding to the inflection points a and b does not change). With further evaporation a heterogenous desaturation starts to develop from the top surface towards the inner part of the sample. To better understand this process, let us consider the different inflection points c to e in the saturation profiles of the large sample (Fig \ref{NMRSmallBig}-A). On the profile at time $t_3$, the relative moisture content around 0,6 is quite homogeneous for heights going from 0 of 29 mm (indicated as point c) and then drops gradually when you approach the free evaporative surface. At $t_4$, the moisture content is equal to roughly 0,4 up to the height 22 mm (marked by point d) and then decreases until the height of 28 mm where the moisture content reaches 0 (green arrow). Thus, at the relative moisture content of 0.4, a dried region of about 2 mm is formed beneath the surface. At $t_5$, at the residual saturation of 0,1, the inflection point e is even further inside the sample, to a height 20 mm and a dry region of 6 mm is present beneath the surface.

Similar results are obtained in small stones, as visible in Fig \ref{NMRSmallBig}-B. At $t_1$, the moisture content in the sample is quite homogeneous and around 0.8 from 0 to 9mm whereas from 9mm to the evaporative surface, the saturation drops down. This transition height, for which the saturation is not homogeneous anymore, moves further inside the sample with time (see points b to d). These drying behaviours are different to what is commonly observed for a simple liquid such as water in porous media (Appendix B). In the case of water, the constant rate evaporation is typically associated with a homogeneous decrease of the overal relative moisture content over the whole height of the sample due to the existence of a continuous liquid film in contact with the free surface of the sample for most of the drying time. This is similar to what we can see at the early stage of the evaporation of the sol (time t1 and t2) in Fig \ref{NMRSmallBig}-A. For water, the CRP ends when the capillary forces are not sufficient anymore to compensate the evaporative flux at the surface and the liquid/air menisci recedes inside the porous material, which causes the formation of a dry receding front in the sample. 

In Fig \ref{NMRcurve}-A and \ref{NMRcurve}-B the drying kinetics obtained from the integration of NMR moisture profiles are plotted and compared with the gravimetric experiments. One can notice that the drying of the sol is much slower than the drying of the pure solvent (water/ethanol) independent of the size of the samples. In large samples, the drying takes $150$ h for the MTS sol against $50$ h for the solvent. For the solvent, the typical constant rate period is followed by a much slower drying that sets in at a very late stage, at almost complete desaturation. For the MTS sol there is no constant rate period but only a falling rate period with two regimes. In the first regime, the variation of the moisture content can be fitted by an exponential decay. In small sample, the end of this first regime coincides with the complete desaturation of the sample whereas for large ones this corresponds to $t_4$ in Fig \ref{NMRSmallBig}-A, for a desaturation of 40$\%$. For the large sample, the second regime can then be fitted by a square-root of time, usually characteristics of diffusion-driven evaporation. The latter coincide well with the change in NMR profiles : the transition between the two regimes corresponds to profile at time $t_4$ (point d) in Fig \ref{NMRSmallBig}-A, profile for which a dry region is forming beneath the surface (see the green arrow). The liquid thus evaporates and has to diffuse inside the stone, leading to a square rort of time kinetics and the development of a growing dry region. It remains to understand the peculiar shape of the saturation profiles in Fig \ref{NMRSmallBig} and why a dry region is forming in the big samples at such high remaining saturation. For that, computed X-ray microtomography which allows to image the fluid is performed.

\subsection{Distribution of MTS determined by computed X-ray microtomography and SEM}
\begin{figure*}[htb]
    \centering
    \includegraphics[width=\textwidth]{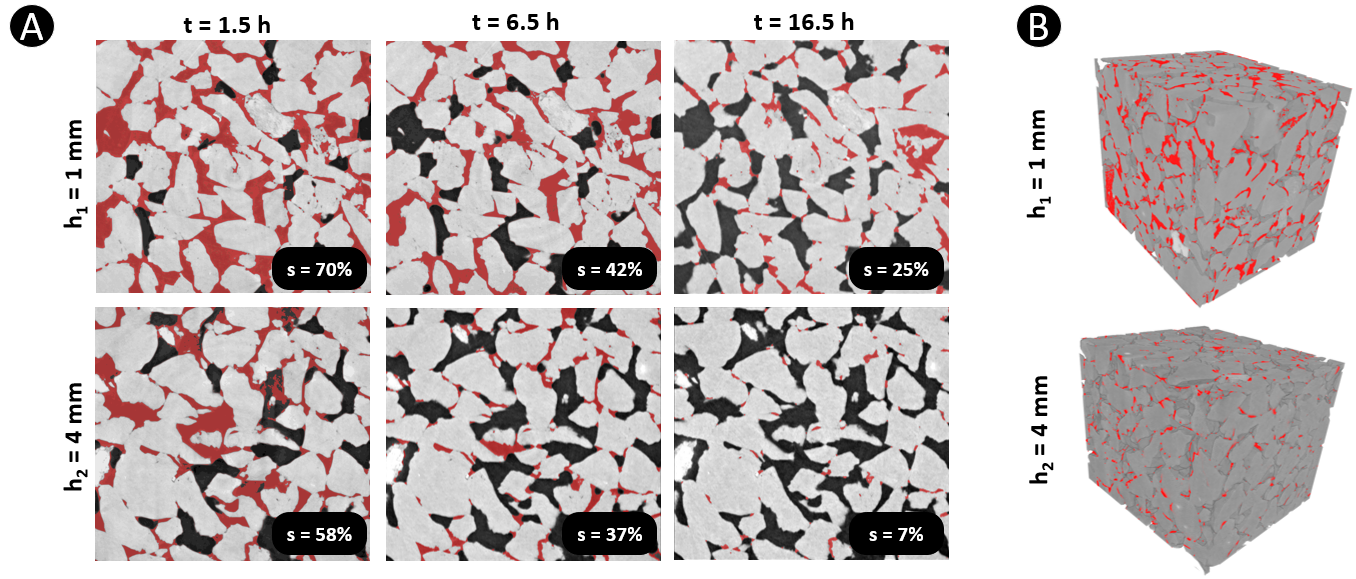}
    \caption{ (A) Slices of scans obtained by computed X-ray microtomography of a stone saturated with sol at different stage of drying. The slice are taken at different times and at depths $h_1=1$ mm and $h_2=4$ mm in the stone. The saturation is calculated over the whole volume of scanned cube. (B) 3D rendering from cubes of volume 0.92 mm$^3$ of the Mšené sandstone treated with MTS at the last stage of drying (the identified gel regions are presented by red color).}
    \label{Tomokinetics}
\end{figure*}

Computed X-ray microtomography is performed at different stages of drying. Fig \ref{Tomokinetics}-A shows the snapshots that display the MTS fluid distribution (represented by red-shaded regions) within stone at 3 different evaporation time from slices at $h_1=1$ mm (top row) or $h_2=4$ mm (bottom row) depth of the stone. The sol or gel state of the liquid cannot be differentiated in the scans. The saturation $s$ values in sol and/or gel (computed from the corresponding 3D volume scans composed of 2000 slices) are indicated for each slice in Fig \ref{Tomokinetics}-A.
The snapshots show that with drying, the larger pores are emptying first followed by the smaller ones, the latter having a higher capillary pressure remains saturated longer  \cite{coussot_scaling_2000, fei_pore-scale_2022}. This is further investigated in Appendix C where the saturation of the pores at different times depending on their diameters is studied. In addition, our results show that the fluid (sol and gel) distribution is not homogeneous inside the stone: at a given evaporation time, there is more fluid (solvent+ macromolecules and/or gel) at the top ($h_1$) than inside ($h_2$) the stone. 
At the end of the drying, the final saturation of gel is $s=25\%$ at the top of the stone whereas $s=7\%$ inside. At both positions the porosity of the stone is reduced by the gel formation. Due to the high saturation of gel at the top, the porosity reduction is more important there, with a porosity at the end of the drying of roughly $\phi =21 \% $ compared to $\phi=28 \%$ for untreated stones. In $h_2$, the final porosity is almost unchanged and equal to $\phi=25 \%$. Fig \ref{Tomokinetics}-B also shows the distribution of gel at $t=16.5$h (corresponding to the end of the drying, see Fig \ref{NMRcurve}-B) in 3D volume scans at the two positions in the stone. Therefore, we can conclude from these measurements that while the gel is mostly formed in small pores, a gradient of gel concentration from the top to the bottom of the sample has developed during drying. 

Comparing the tomography scans in Fig \ref{Tomokinetics}-A (t=6.5h) with the NMR results in Fig \ref{NMRSmallBig}-B, one can notice that whereas the saturation in fluids is higher at the top of the stone in the tomography scans, the moisture content is lower at the top in the NMR measurements. Those two observations can first seem contradictory, but the explanation lies in the fact that with micro-CT, we detect the fluid which can be considered as the solvent with macromolecules and/or potentially the gel whereas NMR data gives only the solvent profiles (\textit{i.e.} the hydrogens in the gel or macromolecules are not detected by the NMR set up). By combining the results of both techniques we can assume that the fluid is thus not homogeneous towards the surface: it is more concentrated in MTS macromelocules and/or gel at the evaporation surface with less solvent. The gelation consequently starts from the top surface in small pores as they remain saturated with fluid longer but with a lower saturation in water and ethanol as detected by NMR. This hypothesis will be further substantiated in the discussion.

\begin{figure}[h!]
    \centering
    \includegraphics[width=0.99\columnwidth]{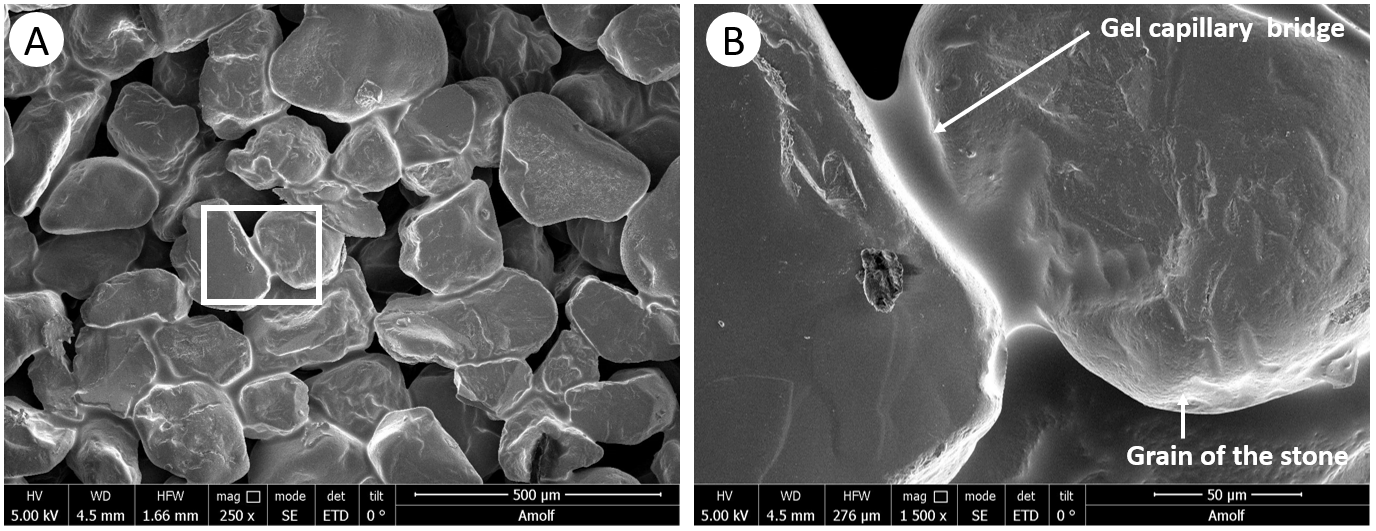}
    \caption{SEM pictures of sandstone grains consolidated with MTS gel}
    \label{sem}
\end{figure}

The consolidated sandstones are also investigated using scanning electron microscopy (SEM) as shown in Fig \ref{sem}.
We observe the formation of a crack-free gel consisting in a thin film which surrounds the grains and connects them together by capillary bridges. In agreement with previous reports on consolidation of stones with TEOS \cite{wheeler_alkoxysilanes_2005,artesani_recent_2020,tsakalof_assessment_2007}, those thin films ensure an efficient consolidation of the material \cite{mosquera_stress_2003, mosquera_producing_2003}. The formation of a crack-free gel within the pore structure with the MTEOS can be explained by the fact that MTS creates a hydrophobic silica network because of the methyl group which in this case is poorly wetted by the polar solvent. This results in lower stresses in the gel network during condensation (lower capillary pressure) and the formation of a gel without cracks \cite{brinker_sol-gel_2013, land_processing_2001}. \\

\subsection{Discussion on drying kinematics during sol-gel transition in porous media } \label{sec:discussion}
\begin{figure*}[htb]
    \centering
    \includegraphics[width=0.99\textwidth]{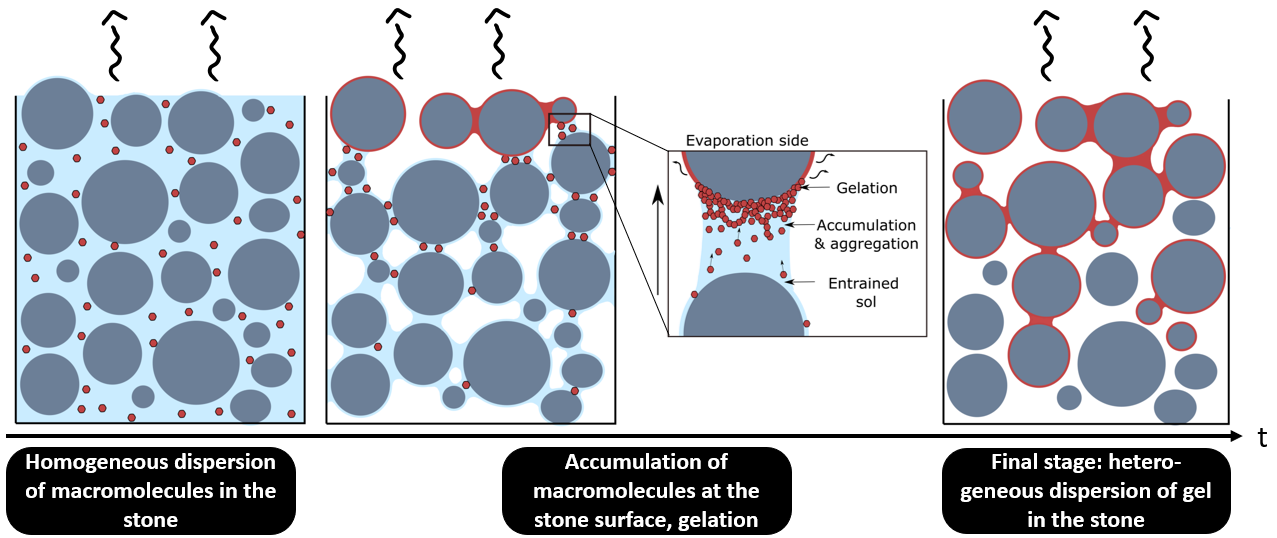}
    \caption{Schematic drawings representing the drying and gel formation in a porous medium.}
    \label{drawings}
\end{figure*}

In this subsection, building on our insights from NMR, X-ray microtomography, gravimetric and SEM experiments, we provide an explanation of drying of the sol in porous media leading to a heterogeneous distribution of gel within the pore network. 

The temporal evolution of the moisture profiles in the samples obtained by NMR technique (Fig \ref{NMRSmallBig}) has shown the development of heterogenenous region from the free surface beneath the surface with the evaporation of the solvent. This results in the decrease of the evaporation rate with time. To better understand the mechanism behind this behaviour, one needs to compare the velocity of liquid transfer to the evaporative surface to the evaporation rate in our experimental conditions. The velocity $v_{Darcy}$ of the liquid transfers in a porous network under a pressure gradient $\frac{\mathrm dp }{\mathrm dx}$ can be approximated by Darcy law: 
\begin{equation}
    v_{Darcy}=-\frac{k}{\mu}\frac{\mathrm dp }{\mathrm dx} 
    \label{darcy}
\end{equation}
where $\mu$ is the viscosity of the liquid and $k$ the permeability of the network. At the beginning of the drying, the porous media is fully saturated and the permeability of the porous network can be determined from an adapted form of the Carman–Kozeny equation for sandstones \cite{rabbani_specific_2014}:
\begin{equation}
    k = \frac{w^3}{17.87*(1-w^2)} d^2
\end{equation}
where $d=70 \mu$m is the mean grain radius of Mšené sandstones. In our situation, the pressure gradient is caused by the heterogeneity of the capillary pressure distribution in the pores network and from Eq.\ref{darcy} the velocity $v_{cap}$ of the capillary transfers in the porous network \cite{coussot_scaling_2000}, assuming that initially the fluid is perfectly wetting the stone is: 
\begin{equation}
    v_{cap}= \frac{k}{\mu (t)} \frac{4 \sigma}{ h r}
    \label{vcap}
\end{equation}
where $\sigma$ and $\rho$ are respectively the surface tension and the density of the sol at the start of the drying. $h$ is the height of the stone and $r$ is the pores radius. Even if during the sol-gel transition, the physical properties of the fluid, {\it e.g.} viscosity evolve (see Appendix A), one can estimate that at the initial stage, the viscosity of the fluid is described by Einstein law's of viscosity $\mu=\mu_0(1+ a \phi)$, $\phi$ being the volume fraction of solute particles (here MTS macromolecules) \cite{mueller_rheology_2009, sacks_rheological_1987, vogelsberger_model_1992}. With $\mu (t=0) = 5*10^{-3}$ Pa.s and $\sigma=28$ mN.m$^{-1}$ and using Eq.~\eqref{vcap} one can thus approximate $v_{cap} (t =0) \approx 10$ m.s$^{-1}$.  On the other hand, the drying rate in a porous medium is calculated as:
\begin{equation}
    v= - \frac{1}{\rho A} \frac{\mathrm{d}m}{\mathrm{d}t}
    \label{evaporationspeed}
\end{equation}
where $m$ is the mass of the liquid in the porous media and $A$ the evaporative surface. Using the experimental results for $\frac{\mathrm{d}m}{\mathrm{d}t}$, this gives $v \approx 10^{-7}$ m.s$^{-1}$ which is much lower than $v_{cap}$.Therefore, the heterogenous decrease of the saturation profiles in Fig \ref{NMRSmallBig} and the formation of the dry region beneath the surface can not be explained by weak capillary forces. 

Originally, the sol imbibed in the stones is a colloidal suspension composed of macromolecules of size $2 - 4$nm as stated before (see Methods - A). With the capillary flows, the macromolecules which are originally homogeneously dispersed in the stone will be advected to the surface as evaporation proceeds, in the same way as ions or solid colloidal particles \cite{keita_mri_2013, keita_particle_2021, lidon_dynamics_2014, qin_lattice_2023}. The calculation of the Peclet number gives an estimation of the order of magnitude of the advective and diffusive transport. The advection of macromolecules to the surface is generally counterbalanced by the diffusion process which tends to equilibrate and homogenize the concentration gradient inside the fluid: 
\begin{equation}
    Pe = \frac{h v}{D} 
\end{equation}
where $v$ is the mean speed of the flow of drying fluid calculated via Eq.~\eqref{evaporationspeed} and $D$ the diffusion coefficient of the MTS macromolecules in the sol. The latter can be deduced from Stokes-Einstein law \cite{miller_stokes-einstein_1997} and is given by: 
\begin{equation}
    D = \frac{kT}{6\pi r \mu (t) }
    \label{diffusion}
\end{equation}
with $r\approx3$ nm the radius of the macromolecules. At the early stage of drying, this leads to $Pe \approx 10^{2}$ over the height $h$ of the stone showing that the diffusion is therefore not sufficient to counterbalance the advection of the macromolecules to the surface. 

The accumulation of macromolecules at the evaporative surface and their depletion from the rest of the stone sample lead to the sol-gel transition in small pores at the surface and a subsequent increase of viscosity there (see Appendix A). In addition, the increase of the viscosity by its own leads also to the decrease of the mean speed of capillary transfer near the evaporating surface according to Eq. \ref{vcap} provided that the change of surface tension during sol-gel transition is negligible (which is indeed the case, see Appendix A). Thus, after the sol-gel transition locally at the microscale, these regions become unavailable for further wetting by the remaining fluid to maintain the liquid front at the whole evaporation surface. As a result, the evaporation rate decreases with time and with the increase of gel regions.

The exponential behaviour in the first drying regime in Fig \ref{NMRcurve} is similar to what was reported by ~\cite{desarnaud_drying_2015} for the evaporation of salt solution when salt crystals cover gradually the evaporative surface of the stone. As explained by \cite{desarnaud_drying_2015}, from Fick's first law, the loss of mass can be related to the evaporative surface: 
\begin{equation}
    \frac{\mathrm{d}m}{\mathrm{d}t} = - D_v A \frac{\rho_{sat}-\rho_{vap}}{\delta} , 
\end{equation}
where $D_v$ is the vapor molecular diffusion coefficient, $\rho_{sat}$ is the vapor density at saturation, $\rho_{vap}$ is the vapor density and $\delta$ is a characteristic transport length over which the relative humidity goes from unity to the atmospheric one. On the other hand, the gradual decrease of the available evaporative surface is proportional to the rate of phase change of mass of fluid (in our case gel formation, in the case of \cite{desarnaud_drying_2015}, salt precipitation):
\begin{equation}
    \frac{\mathrm{d}A}{\mathrm{d}t}= \alpha \frac{\mathrm{d}m}{\mathrm{d}t}
\end{equation}
Combining these two equation leads to : 
\begin{equation}
        \frac{\mathrm{d^2}m}{\mathrm{d}t^2}= - \alpha D_v \frac{\rho_{sat}-\rho_{vap}}{\delta} \frac{\mathrm{d}m}{\mathrm{d}t}
\end{equation}
which has an exponential solution. In samples of small size (volume to surface ratio 10 mm), the end of the exponential decay regime of drying coincides with a complete desaturation of the sample. However, for large stones, this exponential decay regime is followed by a square root of time decrease, as seen in Fig \ref{NMRSmallBig}, characteristic of diffusion limited evaporation (see Appendix B or \cite{coussot_scaling_2000, prat_recent_2002}). This is consistent with our results of NMR and X-ray microtomography where the transition to the diffusive regime coincides with to the saturation profile at $t_4$ in Fig \ref{NMRSmallBig}-A and Fig \ref{NMRcurve}-A for which the first 2mm beneath the surface are dry. The dry region afterward progresses gradually deeper within the stone. A schematic summarizing different stages of the drying is presented in Fig \ref{drawings}. As the stone desaturates, the sol also occupies a smaller fraction of the porosity. Therefore the macromolecules accumulate further away from the free surface and in a smaller fraction of the pores leading to a gradient of gel concentration within the porous network.

\section{Concluding remarks}
To summarize, MTS suspensions can be used to consolidate sandstones via the formation of a crack-free gel within the porous structure. This is of high relevance for cultural heritage applications where damaged stone artworks need to be consolidated. The dynamics of the evaporation of a Newtonian fluid (sol) which becomes gradually a non-Newtonian gel have been investigated using different highly advanced techniques. Combining mass measurements and NMR experiments, we showed that from the initial stage of evaporation, the drying rate continuously and rapidly decreases. Two regimes can be identified: a first regime with an exponential decay and a second diffusive regime. In contrast with the drying of water in porous media, no constant rate period is observed. Computed X-ray microtomography revealed that the concentration of gel formed within the porous network is heterogeneous, contrary to what is believed with such treatment in terms of conservation treatment. A gradient of gel is forming with the highest concentration at the top surface where evaporation takes place. Our results show that gel is found mainly as capillary bridges between grains in the small pores where the capillary pressure is the highest.

The heterogeneous gel distribution can be quantitatively explained by the advection of the macromolecules at the evaporative surface by capillary flows. The accumulation of macromolecules at the surface will lead to the localised sol-gel transition and increase in viscosity. The gel region being hydrophobic, the highly viscous fluid will continue to dry beneath the gel region, leading to the formation of a gradient in solvent saturation and finally the formation of a receding front which will progress inside the porous media with the continuous sol-gel transition. 
In the context of conservation of cultural heritage, our results show that using sol-gel process in heterogeneous porous materials with different pore size distribution, remains more adequate that in homogeneous materials, as the gel will be formed in small pores and the large pores remain empty. In this way, the material although consolidated can still breath thanks to larger pores. On the contrary such treatment in homogeneous pore size distribution materials should induce the skin effect with the gel being concentrated at the subsurface in most of the pores and reducing the permeability as observed by some of the authors in \cite{Solgel2DPM}.

\begin{acknowledgments}
We thank Elham Mirzahossein for the rheology datas and the insightful discussions, Gertjan Bon from the Technology Center of the University of Amsterdam who cut the stones samples and Paul Kolpakov for the help with the SEM pictures. The acquisition of the scans at Synchroton SOLEIL was also only made possible thanks to the help of Jop van der Laan and Valentina Gatto. The authors thank Synchrotron SOLEIL for the ANATOMIX beamline time (proposal n° 20210437). ANATOMIX is an Equipment of Excellence (EQUIPEX) funded by the \textit{Investments for the Future program} of the French National Research Agency (ANR), project \textit{NanoimagesX}, grant no. ANR-11-EQPX-0031. Tinhinane Chekai acknowledges the support from the European project CRYSTINART through the joint Programming initiative on cultural Heritage (JPI-CH). H. Derluyn acknowledges the support from the European Research Council (ERC) under the European Union’s Horizon 2020 research and innovation programme (grant agreement no. 850853).
\end{acknowledgments}

\section*{APPENDIX}

\subsection{Evolution of MTS properties during sol-gel transition}
\begin{figure}[h!]
    \centering
    \includegraphics[width=0.99\columnwidth]{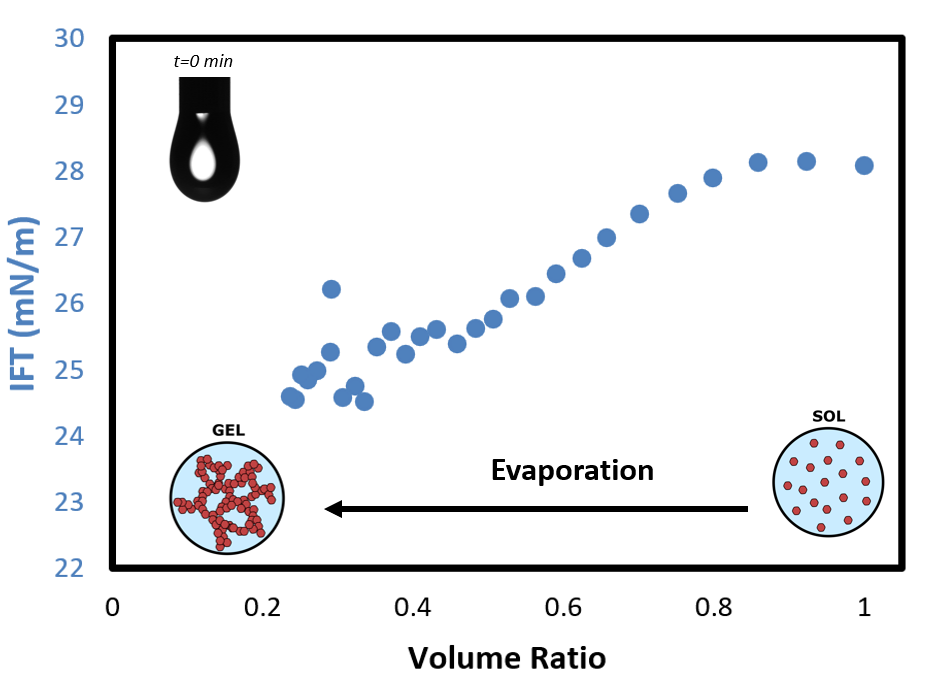}
    \caption{Surface tension evolution during the sol-gel transition of MTS plotted as a function of the volume ratio defined here as $V/V_0$ where $V_0$ is the initial volume of sol}
    \label{surfacetension}
\end{figure}
During the evaporation and sol-gel transition of MTS, the properties of the fluid such as its surface tension and its viscosity evolve. The variation of the surface tension is measured with a Drop Shape Analyser from Kruss using the pendant drop method during the gelation. The results are reported on Fig \ref{surfacetension}. As the gelation proceeds, a decreases in the the surface tension from $28$ mN.m$^{-1}$ to $24$ mN.m$^{-1}$ is observed. 

\begin{figure*}
    \centering
    \includegraphics[width=0.99\textwidth]{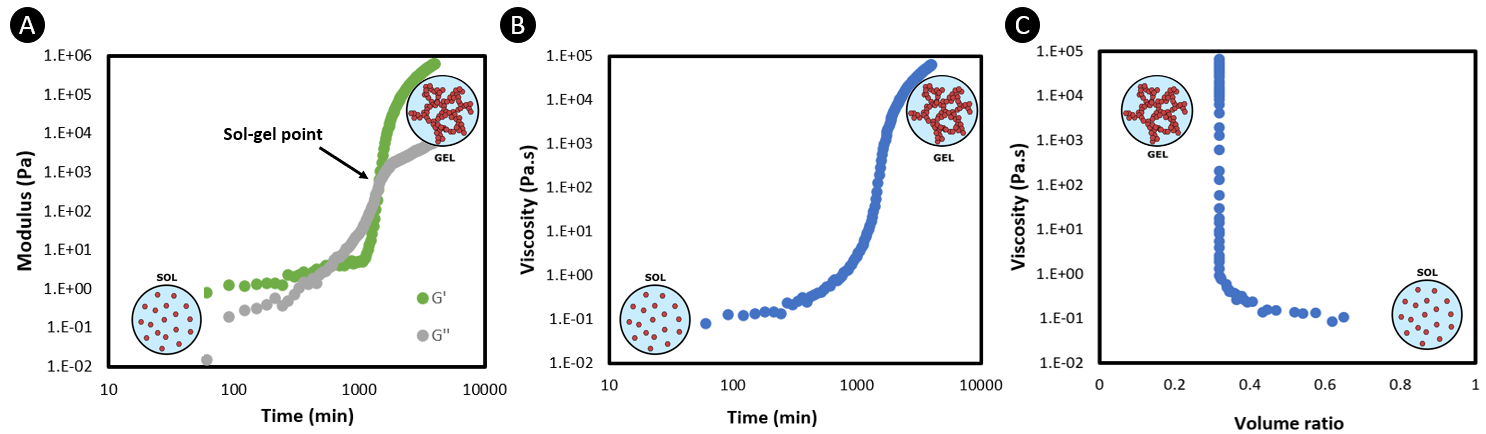}
    \caption{Rheological measurements during the sol-gel transition of MTS. (A) Evolution of the elastic modulus (G') and the viscous modulus (G'') over time. (B) Evolution of the viscosity $\eta = \frac{\sqrt{G'^2 + G''^2}}{\omega}$ over time. (C) Evolution of the viscosity as a function of the volume ratio $V/V_0$ where $V_0$ is the initial volume of sol.}
    \label{viscosity}
\end{figure*}

As explained in section A in Methods, the sol-gel transition is characterized by an important change in the viscoelastic properties of the fluid. Here, we performed rheological measurements during the sol-gel transition of MTS with a rheometer head (Anton Paar, DSR502). The rheology measurement is performed using a 25mm diameter top plate that together with the bottom microscope slide constitutes a plate-plate geometry. The oscillatory shear experiments are done at a frequency of $10$ rad.s$^{-1}$ with a small strain amplitude of 0.25\% to measure the linear viscoelastic response; it was verified that neither the frequency nor the amplitude alter the response. The evolution of the elastic modulus (G') and the viscous modulus (G'') during evaporation are plotted in Fig \ref{viscosity}-A. The sharp increase of the storage modulus G' at $t \approx 1000$ min reflects the formation of clusters and percolation of the macromolecules in the sample. This process carries on until the elastic modulus becomes equal to the viscous modulus, defining the sol-gel point ($t \approx 1440$ min in Fig \ref{viscosity}) \cite{brinker_sol-gel_2013, tung_relationship_1982}. From the measurement of the loss and storage modulus it is possible to deduce the viscosity of the fluid: $ \eta = \frac{\sqrt{G'^2 + G''^2}}{\omega}$ ($\omega$ being the frequency of oscillation) which is plotted in Fig \ref{viscosity}-B.
The viscosity strongly increases during the transition for sol to gel, from $10^{-1}$ Pa.s to $10^{5}$ Pa.s. From Fig \ref{viscosity}-A and \ref{viscosity}-B, we estimate that the sol-gel transition occurs at $t=1440$ min in this configuration which corresponds to a viscosity of $\eta = 80$ Pa.s. In Fig \ref{viscosity}-C, the viscosity increase is plotted as a function of the volume ratio $V/V_{0}$ ($V_0$ being the initial volume of sol) in order to relate to the surface tension evolution. After the sol-gel transition, the volume loss slows down drastically which make it difficult to measure, while the viscosity still increases.

Fig \ref{surfacetension} and Fig \ref{viscosity}-C confirm that in Eq \ref{vcap} and \ref{diffusion} the variation of the surface tension of the fluid can be neglected compared to the variation of the viscosity.\\

\subsection{Drying of water followed by NMR}

\begin{figure*}[htb]
    \centering
    \includegraphics[width=0.99\textwidth]{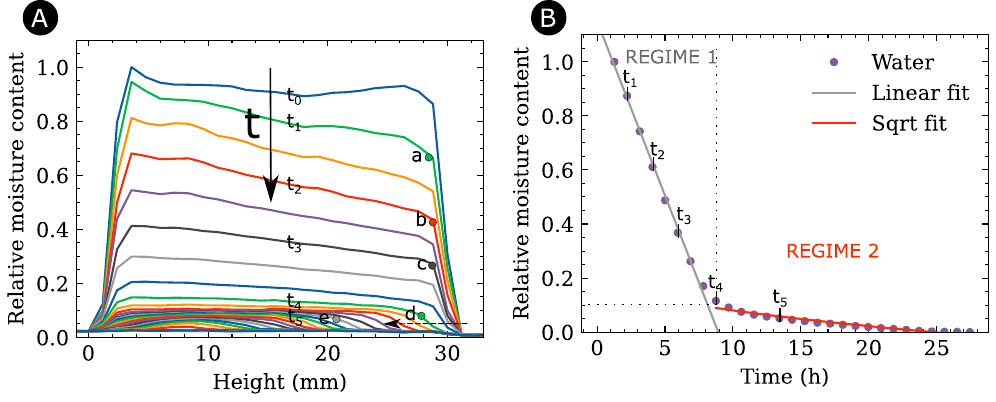}
    \caption{(A). NMR profiles during unidirectional drying of water in a big sandstone. Each profile is afterward integrated over stone height to determine the relative moisture content as a function of time, which is plotted in panel (B). }
    \label{NMRWater}
\end{figure*}
Fig \ref{NMRWater} reports the results for water drying in a sandstone of size $30$mm (heigh) and $20$mm (diameter): \ref{NMRWater}-A shows the moisture distribution as a function of the position of the sample (the top of the stone is at height 2cm) whereas \ref{NMRWater}-B shows the evolution of the total moisture content as a function of time. This is obtained by integrating the profile from \ref{NMRWater}-A. 
In Fig \ref{NMRWater}-A two drying regimes can be identified: a first period (from $t_0$ to $t_4$)where the desaturation occurs homogeneously although the presence of airflow can create a gradient between the top and the bottom slightly tilted curve between top surface and the bottom). A second period (from $t_4$) where the top of the sample gradually dries while the saturation at the bottom remains constant and leads to the apparition of a front between the dry and the partially saturated part (dotted arrow). Those two regimes are also visible in Fig \ref{NMRWater}-B where a sudden decrease in the drying rate is visible at $t_4$ between the first and the second period. 

Indeed, in general, it is well established that for simple Newtonian fluids, two main drying regimes can be observed \cite{coussot_scaling_2000, xu_dynamics_2008, thiery_drying_2017, prat_recent_2002}: the Constant Rate Period (CRP) where evaporation happens at the surface of the stone and the Falling Rate Period (FRP). During the CRP, the capillary transport to the surface of the porous medium is faster than the mean evaporation speed. The water evaporating at the surface is constantly replaced with water from the inner part of porous medium inducing a homogeneous desaturation at a constant rate. It is commonly accepted that the CRP ends when the capillary forces are not sufficient anymore to compensate the evaporative flux at the surface and the liquid/air menisci recedes inside the porous material. The latter causes the formation of a dry front within the porous material and because diffusion will subsequently control the evaporation, the rate will drop in this Falling Rate Period.

\subsection{Evolution of pores saturation during the drying investigated with X-ray microtomography}
From the slices in Fig \ref{Tomokinetics}-A, it is visible that, during evaporation, the big pores are emptying whereas the smaller ones remain saturated. At the end of the drying, the gel is only formed as capillary bridges between grains (Fig \ref{sem}) or in the small pores. 
This observation is substantiated by further analysis on the microtomography scans at position $h_1$. 

\begin{figure}[htb]
    \centering
    \includegraphics[width=0.99\columnwidth]{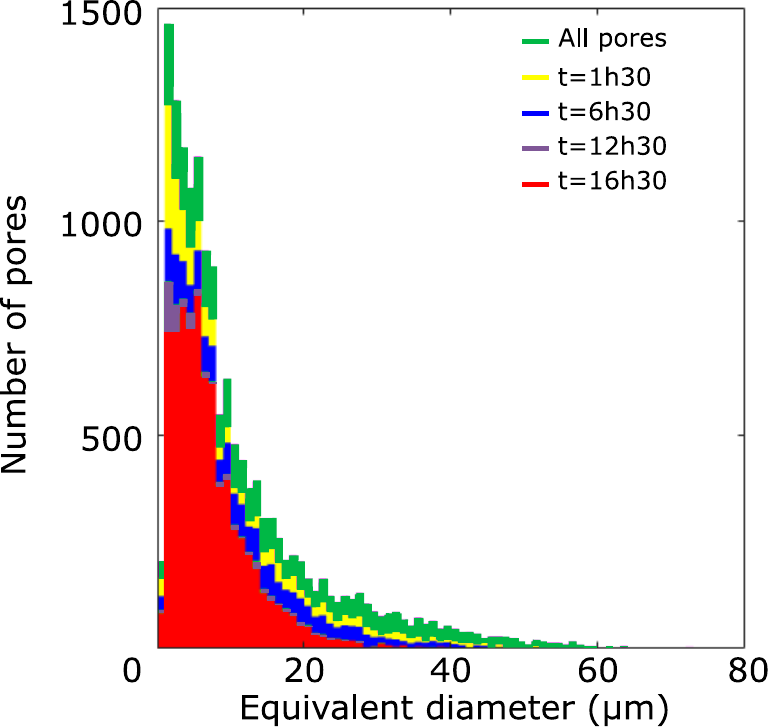}
    \caption{ Histogramms of the pores saturation as a function of the pores diameter at different times. The results are extracted from the microtomography scans taken at position $h_1 =1$ mm.}
    \label{PoresEmptying}
\end{figure}

A pore network is constructed from the scans using the implemented plugin of OpenPNM \cite{gostick_openpnm_2016} within Dragonfly (ORS) software. The OpenPNM plugin is using PoreSpy’s SNOW (Sub-Network of an Over-segmented Watershed) network extraction algorithm. More explanations on the extraction algorithm and the resulting pore network can be found in \cite{these_tinhinane}. Using the aforementioned algorithm and package, we can retrieve properties of each individual pores, the pore size being described by the equivalent diameter. The saturation $s$ (sol and/or gel) of the pores as a function of their diameter is determined at different times of the drying. The results are plotted in Fig \ref{PoresEmptying} with a bin width of 1 $\mu$m.

At $t=1.5$ h, the number of saturated pores has decreased for every diameters and the biggest pores (50 to 60 $\mu$m) are already empty. Similar observations can be done at $t=6.5$ h and $t=12.5$ h.  
At $t=16.5$ h (end of the experiment), only the smallest pores remain saturated: most of the pores with a diameter higher that 25 $\mu$m are empty. Those histograms confirm what was already visible in Fig \ref{Tomokinetics}-A for $h_1$ where at $t=16.5$ h the gel is formed in the smallest pores. This result is in agreement with previous works, as it is now well studied that during evaporation in porous media the smaller pores remain saturated for a longer time with capillary flows (see \cite{fei_pore-scale_2022, coussot_scaling_2000} for example).

\bibliography{references}

\end{document}